# Comparison of metadata with relevance for bibliometrics between Microsoft Academic Graph and OpenAlex until 2020

Thomas Scheidsteger*, Robin Haunschild*

*  *t.scheidsteger@fkf.mpg.de*; r.haunschild@fkf.mpg.de
IVS-CPT, Max Planck Institute for Solid State Research, Heisenbergstr. 1, Stuttgart, D-70569 (Germany)

**Introduction**
Since its launch in 2015, Microsoft Academic Graph (MAG; Sinha et al., 2015) had been a promising new data source for bibliometric analyses due to its large coverage and set of available metadata (Harzing & Alakangas, 2017). Therefore, MAG has been the object of many studies, in particular comparisons with other important bibliographic databases. In one of the last and largest ones, Visser, van Eck, and Waltman (2021) compared MAG with Web of Science, Scopus, Dimensions, and Crossref.
In May 2021, it was announced by the Microsoft Blog (2021) that the Microsoft Academic website, application programming interfaces, and snapshots would retire on December 31, 2021. Soon after that, the non-profit organization OurResearch, aiming at providing "a fully open catalog of the global research system" (OurResearch, 2021), announced they would preserve and incorporate the last full MAG corpus, only excluding patent data, and to continue and hopefully improve it. Another main source of data should be Crossref. In January 2022, OpenAlex (http://docs.openalex.org) was launched and provided API access to their services as well as data dumps for any purposes. The Curtin University's Open Knowledge Initiative (COKI) has already started to monitor the development of OpenAlex, in particular assessing and comparing the value added by OpenAlex to MAG and to Crossref, both in coverage of publications and other research output (Kramer, 2022).
Scheidsteger, Haunschild, Hug, and Bornmann (2018) studied the possibility of using MAG data for the calculation of field- and time-normalized scores. They compared the scores derived from fields of study and coverage in MAG to those derived from subject categories and coverage in Web of Science (WoS). In the present study, we are interested in comparing metadata that are relevant for bibliometric analyses (in particular field and time normalization of citations) of MAG and OpenAlex:
- the coverage of documents over the years,
- the agreement of bibliographic data,
- the numbers of references of each document,
- the kind and distribution of document types,
- the distribution and relation of subject classifications.

**Data and Methods**
*Microsoft Academic Graph (MAG)*
We downloaded the Microsoft Academic Graph (MAG) data set via the Microsoft Azure portal at the end of December 2021 and received data timestamped with 6 December 2021 (Sinha et al., 2015; Tang et al., 2008). We were not able to get newer data at the beginning of 2022 after the official expiration date of the MAG service. According to the OpenAlex Migration Guide (OpenAlex, 2021), no patents have been transferred from MAG to OpenAlex. Therefore, we excluded all items with document type *Patent* from the comparison. In order to facilitate the distinction between the two databases, we keep the case of the document type names as they are used in both databases. In particular, MAG types are written with capital initials. Because MAG data do not contain the full year 2021, we restricted our analyses to the publication years before 2021. Thus, we considered 197,445,041 papers in MAG.

*OpenAlex*
The OpenAlex data dump was retrieved on 9 February 2022 with an update timestamp of 31 January 2022 on the main table (*works*). Both datasets were imported into and processed in our locally maintained PostgreSQL database at the Max Planck Institute for Solid State Research (Stuttgart, Germany). Before the publication year 2021, we have a total of 198,606,165 works in OpenAlex, of which 96,268,256 possess a DOI.

Documents in MAG and OpenAlex can be linked via a unique ID. OpenAlex like MAG only contains linked references. For most works, there are "Fields of Study" available—called "concepts" in OpenAlex and—only there—all linked to a respective Wikidata ID via the table *concepts*. For more details on the approach and the structure of OpenAlex see Priem, Piwowar, and Orr (2022).

**Results**

*Coverage of publication years in both databases*
Only 777 IDs from MAG are not incorporated in OpenAlex, starting with one item in 1952 and reaching a maximum of 201 in 2020. The document types in MAG of these missing items are about 40% *Journal* and *None*, each, and about 15% *BookChapter*. Over the whole period since 1952, of the 777 MAG IDs, 654 have DOIs, most of them could be found in Crossref. 347 of these DOIs contain the ISBN Bookland prefix "978" or "979" and therefore point to books or book chapters, but only one third of them is assigned to the types *Book* or *BookChapter* in MAG. The number 777 of missing MAG IDs exactly matches the difference between the overall number of MAG papers and 197,444,264 OpenAlex works that have a MAG ID associated with them. Of the DOIs, 23 had been associated with more than one MAG ID and—apart from one—all could be found in OpenAlex.

There are 1,161,901 works indexed in OpenAlex that have *no* corresponding record in MAG, 1,108,176 of them having a DOI in OpenAlex, in particular 1,877 documents before 1800, the first publication year in MAG. In the following, only the documents both databases have in common are going to be investigated.

Figure 1 shows the annual numbers of common documents with and without DOI across the years 1980 until 2020. The unexpected decrease of the total number starting in 2017 is due to the shrinking number of documents without a DOI which in turn is by far dominated by the number of documents with *no* document type assigned.

Figure 1: Numbers of common OpenAlex-MAG documents across the years 1980 to 2020

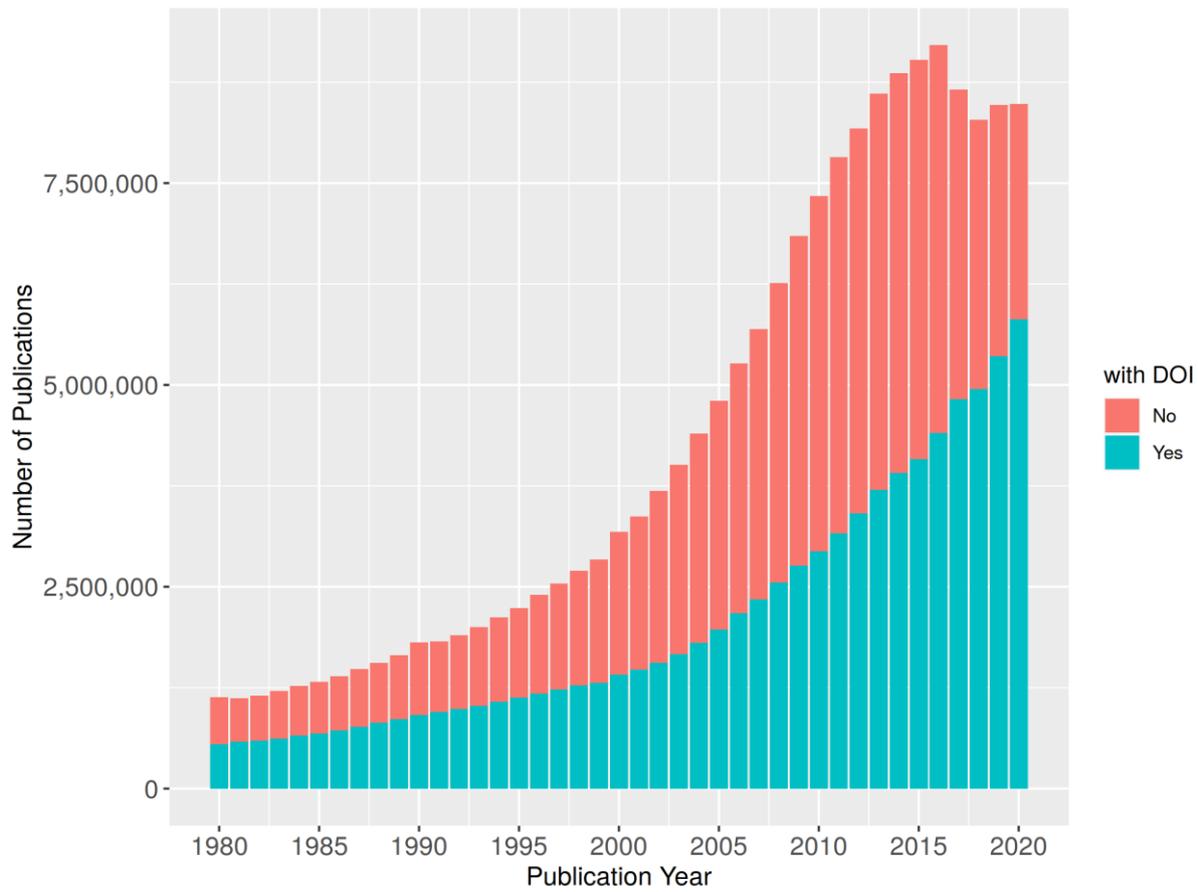

*Comparison of bibliographic data in MAG and OpenAlex*

For the 197,444,264 documents in OpenAlex with an ID in MAG we firstly check if the bibliographic data from MAG, like volume, issue, first page, last page, and DOI are preserved after the transfer to OpenAlex. When volume or issue were available in MAG these data have been completely transferred to OpenAlex. This seems also to be the case for first and last pages and DOIs. But during our investigation we found some issues with the (original MAG) data quality: (i) In more than 28,800 cases, the fields "first page" and "last page" contained not a single number but the same range of numbers, e.g., "35–46". (ii) More than 810,028 DOIs occur *more than once* in the dataset, 7,626 of them at least ten times, and 235 at least 100 times. Of the top 100 most-frequently occurring DOIs, only 29 can be resolved. (iii) More than 6,000 DOIs contain non-latin characters, less than 200 could be resolved. Secondly, concerning the number of (linked) references for a document, we compared the respective values in both databases and found *no* difference.

*Document types in both databases*

In MAG, we are dealing with seven document types: *Book, BookChapter, Conference, Dataset, Journal, Repository,* and *Thesis*. Nearly 45% of the documents are classified as *Journal*, but nearly the same number of documents have *no* document type assigned (*None*).

Table 1: Number and percentages of document types in MAG.

| Document types in MAG | Number of items | Percentage of items |
|---|---:|---:|
| *Journal* | 87,430,385 | 44.28 |
| *None* | 85,844,335 | 43.48 |
| *Thesis* | 5,925,439 | 3.00 |
| *Conference* | 5,053,232 | 2.56 |
| *Repository* | 4,779,269 | 2.42 |
| *Book* | 4,588,285 | 2.32 |
| *BookChapter* | 3,691,552 | 1.87 |
| *Dataset* | 132,544 | 0.07 |
| Sum | 197,445,041 | 100.00 |

In OpenAlex, there are 26 document types that inherit their definition from another major data source Crossref – as documented in Crossref's Content Type Markup Guide (Crossref, 2021). Obviously, all works in OpenAlex with a Crossref DOI receive their document type from there. Those document types with a share of more than 1.0% of all documents are listed in Table 2. There are additional nine million items in OpenAlex assigned to the document type *journal-article* as compared to the MAG document type *Journal*. The OpenAlex items of document type *journal-article* cover nearly one half of all documents, but the items without a document type (*none*) are still more than a third of all. However, the document types *Journal* and *None* are about equally large in MAG. The increased numbers of journal articles, conference proceedings and book chapters are especially interesting from a bibliometric point of view.

Table 2: Numbers and percentages of document types in OpenAlex.

| Document types in OpenAlex | Number of items | Percentage of items |
|---|---:|---:|
| *journal-article* | 96,547,138 | 48.61 |
| *none* | 70,155,602 | 35.32 |
| *book-chapter* | 9,588,895 | 4.83 |
| *proceedings-article* | 7,051,207 | 3.55 |
| *dissertation* | 6,126,640 | 3.08 |
| *book* | 4,522,989 | 2.28 |
| *posted-content* | 3,093,874 | 1.56 |
| other types | 1,519,820 | 0.77 |
| sum | 198,606,165 | 100.00 |

As displayed in Table 3, about 90.1% of all items have the obviously equivalent document types in both databases.

Table 3: Shares of transfers of equivalent document types between MAG and OpenAlex

| MAG | OpenAlex | Number of items | Percentage of items |
|---|---|---:|---:|
| *Journal* | *journal-article* | 86,395,430 | 43.76 |
| *None* | *none* | 70,154,418 | 35.53 |
| *Thesis* | *dissertation* | 5,917,802 | 3.00 |
| *Book* | *book* | 4,421,867 | 2.24 |
| *Conference* | *proceedings-article* | 4,285,360 | 2.17 |
| *BookChapter* | *book-chapter* | 3,662,705 | 1.86 |
| *Repository* | *posted-content* | 3,018,186 | 1.53 |
| *Dataset* | *dataset* | 132,421 | 0.07 |
| | Sum | 177,988,189 | 90.15 |

The more interesting cases are the reclassifications. Therefore, we show in Figure 2 an alluvial diagram of the corresponding document types in both databases, *excluding* the transfers from Table 3. The alluvial diagram was produced using the software package "alluvial" (Bojanowski & Edwards, 2016) based on R (R Core Team, 2020).

Figure 2: Alluvial diagram of document type reclassifications from MAG to OpenAlex

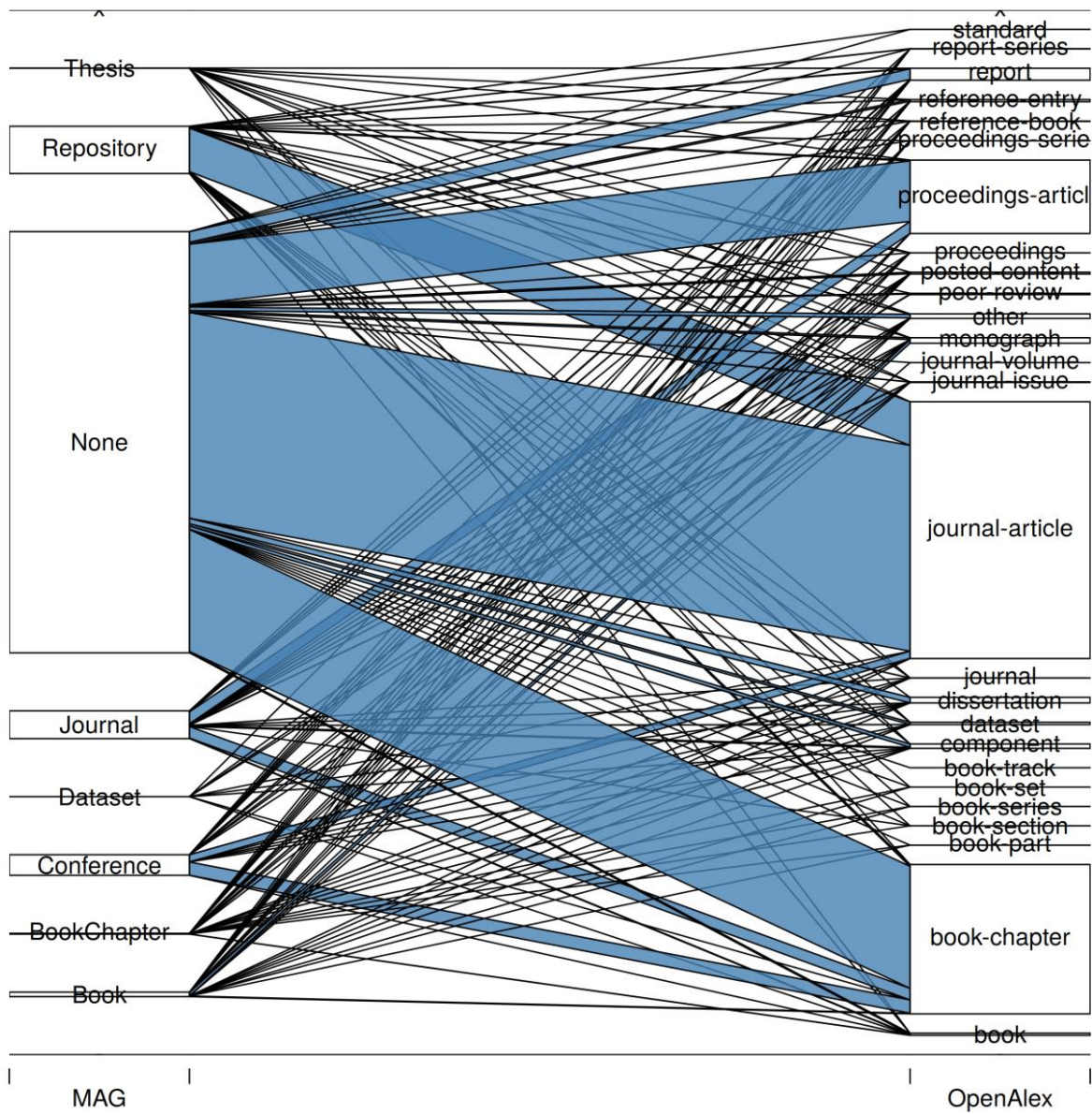

Those reclassifications occurring in relevant numbers that sum up to nearly 9.3% of all documents are listed in Table 4. In order to get an impression of the quality of these reclassifications, we add some characteristics of respective random samples of ten documents, each. All of them had a DOI – as we could expect because of Crossref being the main source of document type information. Indeed, less than 10,000 documents *without* a DOI have been reclassified, i.e. about 0.05% of all 20 million reclassifications.

Table 4: Shares of reclassifications of document types from MAG to OpenAlex together with some characteristics of corresponding random samples of ten documents. Shares of at least 0.1% are shown.

| Document types | | | Random samples of ten documents | |
|---|---|---|---|---|
| MAG | OpenAlex | Percentage of all documents | Span of publication years | Other characteristics |
| *None* | *journal-article* | 3.88% | 1928 - 2018 | 8 titles with Cyrillic, far-eastern, or Arab character set; 1 Dutch document with English title; |
| *None* | *book-chapter* | 2.30% | 1984 - 2020 | All DOIs containing the Bookland prefix "978"; 1 German title |
| *None* | *proceedings-article* | 1.14% | 1971 - 2019 | 7 Cyrillic or Arab titles; only 2 conference papers identifiable |
| *Repository* | *journal-article* | 0.82% | 1988 - 2016 | 4 ChemInform Abstracts; 5 arXiv papers: all DOIs point to published papers; 1 SSRN preprint from 2012, published in 2016 in a journal |
| *Conference* | *book-chapter* | 0.25% | 2001 - 2020 | All published in conference proceedings by Springer as part of a book series; 8 DOIs contain Bookland prefix "978"; 7 documents from LNCS; only one document noted by Springer as chapter, the others as conference papers |
| *Journal* | *proceedings-article* | 0.23% | 2010 - 2017 | 3 poster presentation abstracts in the supplement of a journal; 4 documents from the Proceedings of SPIE; 2 documents in proceedings of a medical conference as supplement to a journal. |
| *Journal* | *book-chapter* | 0.22% | 1965 - 2017 | All book chapters; 8 DOIs contain Bookland prefix "978" |
| *Conference* | *journal-article* | 0.14% | 1987 - 2014 | No conference papers; 4 publishers incorrect in MAG |
| *None* | *report* | 0.20% | 1964 - 2019 | 7 technical reports or geological survey data from US and Canadian government |
| *None* | *dissertation* | 0.10% | 1973 - 2018 | Theses and dissertations at institutional repositories (5 US, 4 Brasilian, 1 Greek) |

The reclassification to type *book-chapter* in OpenAlex seems to work fairly well. This is also the case for *journal-article*. In particular, many documents using non-latin character sets are now getting classified, and a substantial number of items with DOIs that MAG had labelled as arXiv preprints are correctly recognized as *journal-article*. On the other hand, the assignment of ChemInform abstracts to this document type is debatable, but they are definitely no preprints. Conference papers seem to be a special case: Documents incorrectly assigned to *Journal g*et corrected to *proceedings-article*, but for documents without a document type in MAG the assignment of *proceedings-article* is not that accurate or at least difficult to verify. In case of MAG type *Conference*, the reclassification to *journal-article* seems to be overall valid, whereas the reclassification of LNCS contributions to *book-chapter* seems to be the result of their appearance as part of book series and of the format of their DOIs containing the Bookland prefix "978" (doi.org, 2019). This fact should be kept in mind for bibliometric studies in computer sciences, which probably should include book chapters as well.

*Subject Classifications*
OpenAlex states in their migration guide (OpenAlex, 2021) that they use the same taxonomy as MAG but have reduced the number of "Fields of Study" (FoS) by removing those with less than 500 papers associated. Moreover, they have applied a different algorithm, i.e. model V1 in their open-source software (Priem & Piwowar, 2022).

A quick look reveals the persistence of all 19 top-level FoSs (level=0) from MAG as well as of 284 of the 292 FoSs of the next level (level=1). Table 5 lists the distribution of all FoS levels from 0 to 5 in both databases. The strongest reduction of FoS numbers occurs in the levels 3 to 5 where less than 10% persist. The total number of FoSs on all levels is 714,971 in MAG and only 65,073 in OpenAlex, which means a reduction to 9.1%. Interestingly, of the 24,768 level-3 FoSs in OpenAlex, more than 4,000 have less than 500 works assigned to them.

Table 5: Distribution of FoSs in MAG and OpenAlex

| Level | #MAG | #OpenAlex | Difference (#MAG - #OpenAlex) | Percentage (#OpenAlex/#MAG*100.0) |
|---|---|---|---|---|
| 0 | 19 | 19 | 0 | 100.00 |
| 1 | 292 | 284 | 8 | 97.26 |
| 2 | 137,415 | 21,460 | 115,955 | 15.62 |
| 3 | 330,275 | 24,768 | 305,507 | 7.50 |
| 4 | 134,843 | 12,406 | 122,437 | 9.20 |
| 5 | 112,127 | 6,136 | 105,991 | 5.47 |
| All levels | 714,971 | 65,073 | 649,898 | 9.10 |

Even if the top-level FoSs persist, they are very differently associated to the papers. For example, one paper (https://api.openalex.org/works/W2178938397, accessed on 26 April 2022) had one top-level FoS and one level-1 FoS in MAG, but it has six additional top-level FoSs and one additional level-1 FoS in OpenAlex.

The total number of papers with any FoS is significantly increased: 30.5 of 48.9 million documents without any FoS in MAG have at least one FoS in OpenAlex so that the coverage increases from 74.6% to 86.6%. There are 147,360,860 papers with at least one top-level FoS and a total number of 147,426,219 assignments to top-level FoSs in MAG, i.e., 65,359 of the papers have more than one top-level FoS (up to seven). In OpenAlex, there are 170,900,225 works with any top-level FoS, and 229,560,450 assignments to top-level FoSs in total; there are 52,966,153 works with at least two top-level FoSs (up to seven). About 77.2% of all top-level assignments in MAG persist in OpenAlex, but this proportion varies significantly across the 19 top-level FoSs as Table 6 clearly shows – from less than a quarter for Engineering to more than 90% for Material Sciences and Medicine.

Table 6: Distribution of top-level FoSs in MAG and percentage of top-level FoSs persistent in OpenAlex.

| FoS | #MAG | #OpenAlex | % persistent |
| --- | --- | --- | --- |
| Art | 3,717,975 | 2,620,365 | 70.48 |
| Biology | 13,169,649 | 10,411,044 | 79.05 |
| Business | 5,174,422 | 4,200,803 | 81.18 |
| Chemistry | 14,191,693 | 12,194,451 | 85.93 |
| Computer science | 12,312,525 | 10,878,013 | 88.35 |
| Economics | 3,130,346 | 2,131,877 | 68.10 |
| Engineering | 8,472,749 | 2,023,815 | 23.89 |
| Environmental science | 3,533,640 | 2,712,884 | 76.77 |
| Geography | 4,447,923 | 2,366,289 | 53.20 |
| Geology | 3,061,102 | 2,302,537 | 75.22 |
| History | 3,059,007 | 1,650,999 | 53.97 |
| Materials science | 11,063,791 | 10,010,937 | 90.48 |
| Mathematics | 6,021,856 | 4,028,415 | 66.90 |
| Medicine | 27,897,600 | 25,953,084 | 93.03 |
| Philosophy | 2,010,846 | 1,240,834 | 61.71 |
| Physics | 6,873,294 | 5,517,376 | 80.27 |
| Political science | 6,775,718 | 4,899,049 | 72.30 |
| Psychology | 8,063,945 | 6,198,019 | 76.86 |
| Sociology | 4,448,138 | 2,520,233 | 56.66 |
| All documents | 147,426,219 | 113,861,024 | 77.23 |

Figure 3 shows an alluvial plot of the transfer of paper-based subject classifications *without* the persistent FoS assignments of Table 6 so that the remaining reclassifications become more visible. Given the fact that *all* 342 possible reclassifications do indeed occur in our publication set, only the 94 connections with at least 200,000 occurrences are shown. Several reclassifications occur in comparable measures in both directions, e.g., in the pairs Sociology & Psychology, Sociology & Political Science, or Psychology & Medicine. Other ones show a significant transfer in mainly one direction, like Engineering to Computer Science, Mathematics to Computer Science, Biology to Chemistry, or Chemistry to Materials Science.

Figure 3: Alluvial diagram for the top-level FoS reclassifications from MAG to OpenAlex, showing only reclassifications that occur at least 200,000 times

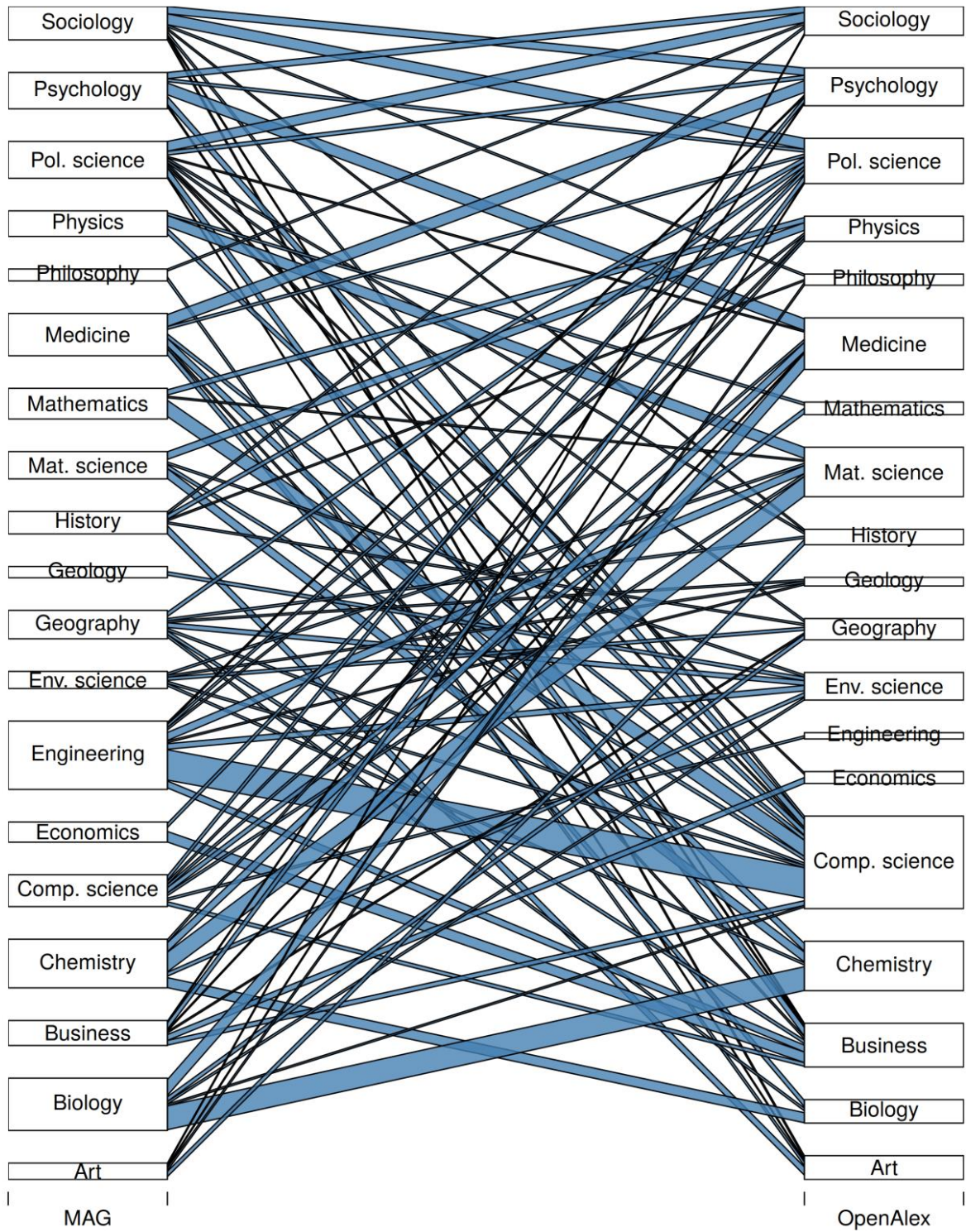

**Discussion and Conclusions**

OpenAlex has transferred practically all works from MAG preserving their bibliographic data publication year, volume, first and last page, DOI as well as the number of references that are important ingredients of citation analysis.

More than 90% of the MAG documents have equivalent document types in OpenAlex. Of the remaining ones, especially reclassifications to the OpenAlex document types *journal-article* and *book-chapter* seem to be valid and amount to more than 7%, so that the document type specifications have improved significantly from MAG to OpenAlex. So far, OpenAlex seems to be more suited for bibliometric analyses than MAG.

As last item of bibliometric relevant metadata, we looked at the paper-based subject classification via FoS in MAG and in OpenAlex. We found significantly more documents with a FoS assignment in OpenAlex than in MAG. On the first and second level, the FoS structure is identical resp. nearly identical, but on the deeper levels the number of available FoSs is drastically reduced to about 10%. But this would not pose a problem if using only the upper two levels for bibliometric analyses as was done by Scheidsteger, et al. (2018). However, the reclassifications might cause changes to conclusions of previous studies. The consequences of the proliferation and abundant reclassification of top-level FoSs need to be studied more in detail. Reclassifications at the deeper levels should be studied, too.

Overall, OpenAlex seems to be at least as suited for bibliometric analyses as MAG for publication years before 2021. However, this first impression needs to be checked by further detailed studies.

**Acknowledgements**

We thank Jason Priem for very helpful comments on an earlier draft.